\documentclass[sigconf]{acmart}
\AtBeginDocument{%
  }

\setcopyright{acmlicensed}
\copyrightyear{2026}
\acmYear{2026}
\acmDOI{XXXXXXX.XXXXXXX}

\usepackage{graphicx}
\usepackage{subcaption}
\usepackage{svg}

\begin{document}

\title{\textsc{StructuredEdit}: Constraint-Aware Graphic Design Editing via Differentiable Parameter Propagation}

\author{Veeramanohar Avudaiappan}
\affiliation{%
  \institution{Department of Electrical and Electronics Engineering, Amrita School of Engineering, Coimbatore, Amrita Vishwa Vidyapeetham}
  \country{India}
}
\email{cb.en.u4eee19145@cb.students.amrita.edu}
\orcid{0009-0003-1227-5394}

\author{Ritwik Murali}
\authornote{Corresponding Author}
\affiliation{%
  \institution{Department of Computer Science and Engineering, Amrita School of Computing, Coimbatore, Amrita Vishwa Vidyapeetham}
  \country{India}
  }
\email{m\_ritwik@cb.amrita.edu}
\orcid{0000-0002-1269-2257}

\renewcommand{\shortauthors}{Avudaiappan \& Murali}

\begin{abstract}
Graphic design editing requires precise manipulation of typography, layout, and visual hierarchy under strict design constraints. Following the introduction of large language models, organizations have increasingly promoted vision–language models to enhance productivity. However, current models operate on pixels and achieve only 52\% constraint satisfaction on structured design edits, thereby limiting their reliability for professional workflows. We present \textsc{StructuredEdit}, a pipeline that reframes design editing as parameter manipulation rather than pixel generation. Our core technical contribution is Differentiable Parameter Propagation (DPP), a training method that embeds hard design constraints into vision-language model fine-tuning by backpropagating pixel-level constraint violations through a lightweight differentiable rasterizer. A hybrid candidate-and-filter pipeline produces 125k validated edit triplets. The resulting system reaches 89\% constraint satisfaction versus 52\% for GPT-4V, 0.82 matched-element IoU, and 76\% top-1 font accuracy over the 100 most-frequent design typefaces. In a user study (N=35), editing time drops 33\% and correction iterations drop 44\% relative to a GPT-4V baseline.
\end{abstract}




\maketitle

\section{Introduction}


Graphic designers spend significant time on routine but fine-grained edits like resizing elements, adjusting colours, and repositioning content. Current AI tools fail at these precise, constraint-governed operations. In fact, frontier Vision-Language Models (VLMs) reach only 23.7\% top-1 font identification across 167 families \cite{Shahgir2025FRB}, and pixel-based editors introduce 29--68\,px positioning errors that violate alignment, hierarchy, and overlap constraints. Existing work addresses this problem along three axes: \emph{1. Layout generation} methods~\cite{Seol2024PosterLlama} that handle placement but not typography or colour, \emph{2. Hierarchical design frameworks} such as COLE~\cite{Jia2023Cole} which produce multi-layer designs from text but lack a natural-language editing interface with constraint guarantees, and \emph{3. Pixel-based editors} ~\cite{Brooks2023InstructPix2Pix} that cannot enforce hard constraints at inference time.

Towards solving this, we present \textsc{StructuredEdit}, which reframes design editing as \emph{parameter manipulation} over a typed layer representation rather than pixel generation. Our core contribution is \textbf{Differentiable Parameter Propagation (DPP)} which involves hard design constraints being embedded directly into VLM fine-tuning by backpropagating pixel-space violations through a differentiable rasterizer. This makes rasterization a training-time supervisory signal. The proposed pipeline, \textsc{StructuredEdit} achieves 89\% constraint satisfaction versus 52\% for GPT-4V, with a 33\% reduction in editing time and 44\% fewer correction iterations in a user study ($N{=}35$).

\section{Methodology}

\begin{figure}
    \centering
    \includegraphics[width=0.85\linewidth]{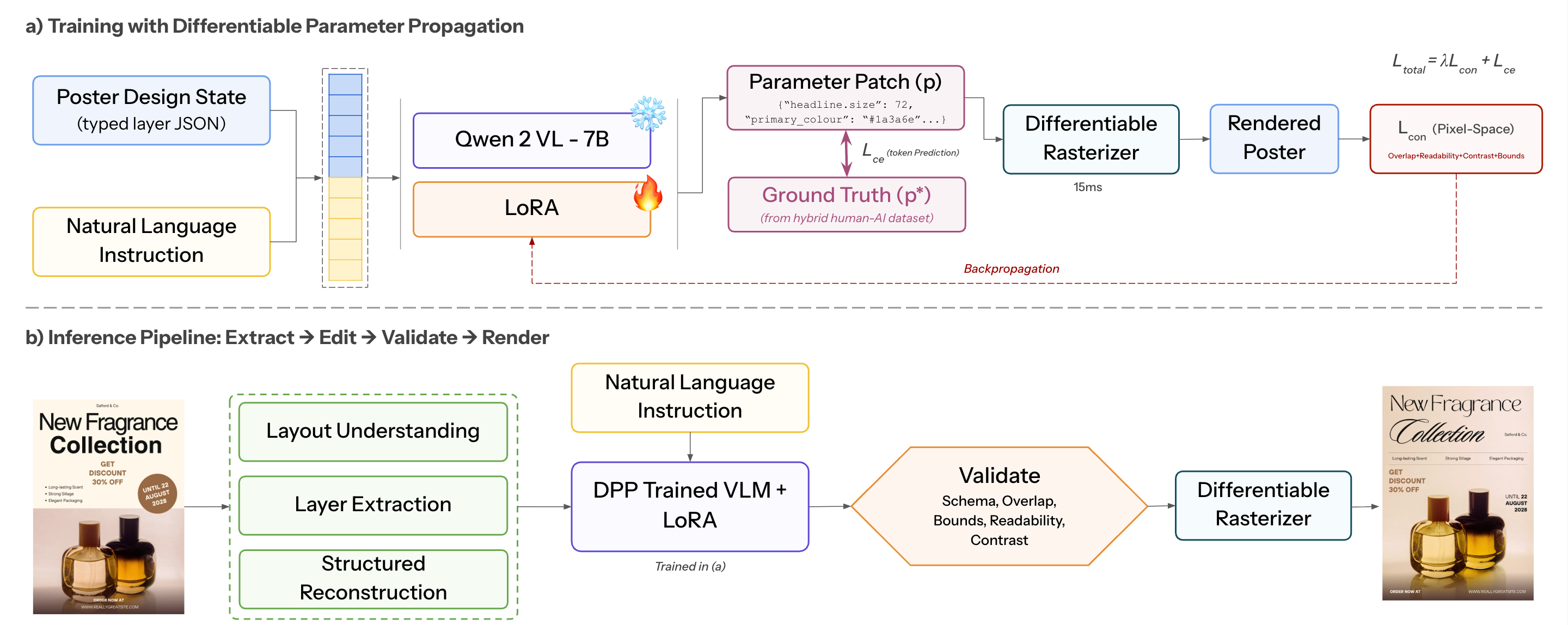}
    \caption{(a)~A LoRA-adapted Qwen2-VL-7B predicts a parameter patch~$p$ from a typed layer JSON and instruction, trained via cross-entropy and pixel-space constraint losses backpropagated through a differentiable rasterizer.
    (b)~At inference, a raster poster is decomposed into structured layers, edited by the DPP-trained model, validated, and rendered.}
    \label{fig:sys-arch}
\end{figure}

StructuredEdit frames graphic design editing as \textit{parameter manipulation over a structured representation} rather than pixel generation. As shown in Figure~\ref{fig:sys-arch}, the system operates in two phases: a training phase that teaches a VLM to produce constraint-satisfying parameter patches via DPP, and an inference pipeline that executes \textsc{Extract} $\rightarrow$ \textsc{Edit} $\rightarrow$ \textsc{Validate} $\rightarrow$ \textsc{Render} on any raster input. 

%
\label{sec:dpp}


\noindent \textbf{(i) Differentiable Parameter Propagation:} As shown in Figure~\ref{fig:sys-arch}(a), a LoRA-adapted Qwen2-VL-7B  takes typed layer JSON and a natural-language instruction as input. It outputs a parameter patch~$p$, supervised against ground-truth~$p^{*}$ using token-level cross-entropy ($\mathcal{L}_{\text{ce}}$), as in standard SFT. Training converges at approximately $3$k steps but plateaus at a $32\%$ constraint violation rate since geometric errors, such as overlaps, low contrast, or out-of-bounds placement, incur no token-level penalty. Post-hoc validation partially mitigates this but yields a $35\%$ rejection rate and provides no learning signal. Consequently, the model fails to generalise to constraint-critical regions it never learned to avoid. DPP addresses this by simultaneously rendering $p$ through a lightweight differentiable rasterizer and backpropagating pixel-space constraint losses ($\mathcal{L}_{\text{con}}$)to the LoRA weights jointly with $\mathcal{L}_{\text{ce}}$. Therefore, the total loss $\mathcal{L}_{\text{total}} = \mathcal{L}_{\text{ce}} + \mathcal{L}_{\text{con}}$ where $\mathcal{L}_{\text{con}}$ is sum of $\mathcal{L}_{\text{overlap}},\mathcal{L}_{\text{read}},         \mathcal{L}_{\text{contrast}},\mathcal{L}_{\text{bounds}}$. $\mathcal{L}_{\text{overlap}}$ is the penalty for pairwise layer IoU~$>$~0.01. Font sizes outside the 8--120\,pt readability range via squared hinge losses are penalized as $\mathcal{L}_{\text{read}}$, WCAG luminance contrast below 4.5:1 as $\mathcal{L}_{\text{contrast}}$, and squared out-of-canvas displacement for all elements as $\mathcal{L}_{\text{bounds}}$, respectively. Since each term is computed directly on rendered pixels rather than on token predictions, the model receives corrective gradients precisely where SFT is blind, for instance, a font size that is numerically valid in one design can violate readability at a different canvas resolution, a dependency that token supervision cannot encode but raster-space loss captures naturally. This shifts training into a second convergence regime beyond the SFT plateau: $\mathcal{L}_{\text{con}}$ continues to drive violation rate from 32\% down to 6\% between 3k and 12k steps without degrading $\mathcal{L}_{\text{ce}}$, confirming that DPP reshapes the constraint-violating region of the output distribution rather than compromising fluent parameter prediction.


 
\label{sec:layers}
 

\noindent \textbf{(ii)Typed Layer Representation:} Each raster input is decomposed into typed layers encoded as structured JSON using RAM for element tagging, Grounding DINO for bounding boxes, Tesseract OCR with a top-100 Google Fonts classifier for text, SAM for image segmentation, and LaMa for background recovery ~\cite{Avudaiappan2025PosterDecomp}; each layer then emits fully editable typed parameters: text exposes \texttt{font\_family}, \texttt{size\_pt}, \texttt{color\_hex}, and \texttt{position}; shapes expose geometry, fill, and opacity; images expose scale and crop; backgrounds expose fill type and gradient, forming the edit surface onto which every DPP constraint loss term directly maps.

\label{sec:validation}
 
\noindent \textbf{(iii)Constraint Validation and Rendering:} Even with DPP training, a deterministic safety pass validates all outputs before delivery. Hard constraints are checked in sequence: (i)~JSON schema validity; (ii)~text overlap (IoU $< 0.01$); (iii)~canvas bounds; (iv)~readability ($8$--$120\,\text{pt}$); and (v)~WCAG contrast ratio $\geq 4.5{:}1$. Residual violations trigger resampling up to five times before returning the identity edit, ensuring that 100\% of delivered outputs satisfy all hard constraints at inference time. All edits are structured as JSON Patch Operations \cite{bryan2013rfc}. Validated patches are rendered by the same differentiable rasterizer used during training, referencing external SVGs for vector layers and raster assets for image layers to preserve export compatibility with downstream design tools.
 
\label{sec:data}
 

\noindent \textbf{(iv)Hybrid Data Curation:} Pure human authoring is costly as it takes  ${\approx} 5\,\text{min}$ per valid edit patch. Purely synthetic generation is inexpensive but unreliable. Here, we adopt a \textit{candidate-and-filter} paradigm. \textit{Qwen2-VL-7B} generates five diverse (instruction, patch) candidates for each design. Professional designers review these candidates and select the best one. Each selection takes $\approx 30\,\text{s}$. Invalid candidates are not discarded but retained as hard negatives. This process produces \textit{$125,000$ validated edit triplets}, derived from $25,000$ templates including $23,651$ from Crello ~\cite{yamaguchi2021canvasvae} and $1,349$ manually curated examples. The dataset includes tasks from five categories. Typography edits(32\%), Layout and repositioning (26\%), Colour and style changes (21\%), Content replacement (13\%), and Hierarchy and scale adjustments (8\%). It  differs from CrelloInstruct ~\cite{Kikuchi2025} which focuses on precise instructions for design completion. In contrast, our instructions are intentionally underspecified. They reflect real-world queries from non-designers.

\begin{table*}[htbp]
\centering
\small
\begin{minipage}[t]{0.62\textwidth}
\centering
\caption{Overall performance and ablation}
\label{tab:main}
\setlength{\tabcolsep}{5pt}
\begin{tabular}{lccccc}
\toprule
Method & Param & Cons. & Font & IoU & \$/1k \\
\midrule
GPT-4V (3-shot)   & 0.34 & 0.52 & 0.28 & 0.51 & 28.50 \\
Standard SFT (Qwen2-VL-7B)                 & 0.74 & 0.68 & 0.63 & 0.73 & 0.58  \\
\midrule
\textbf{StructuredEdit}                    & \textbf{0.84} & \textbf{0.89}
                                           & \textbf{0.76} & \textbf{0.82} & 0.58 \\
\midrule
\quad w/o DPP                              & 0.74 & 0.68 & 0.63 & 0.73 & 0.58 \\
\quad w/o hybrid data                      & 0.72 & 0.81 & 0.68 & 0.75 & 0.58 \\
\quad w/o diff.\ render                    & 0.79 & 0.76 & 0.71 & 0.77 & 0.58 \\
\bottomrule
\end{tabular}
\end{minipage}
\hfill
\begin{minipage}[t]{0.35\textwidth}
\centering
\caption{User study ($N{=}35$, 6 edits each).
$^{\ast}p{<}0.01$ Wilcoxon signed-rank.}
\label{tab:userstudy}
\setlength{\tabcolsep}{4pt}
\begin{tabular}{lcc}
\toprule
Metric & GPT-4V & Ours \\
\midrule
Task time (min)      & 4.6 & 3.1$^{\ast}$ \\
Corrections          & 3.2 & 1.8$^{\ast}$ \\
Satisfaction (7-pt)  & 4.3 & 5.7$^{\ast}$ \\
Trust (7-pt)         & 3.1 & 5.4$^{\ast}$ \\
SUS                  & 58  & 79           \\
\bottomrule
\end{tabular}
\end{minipage}
\end{table*}

\section{Results and Analysis}
\begin{figure}
    \centering
    \includegraphics[width=\linewidth]{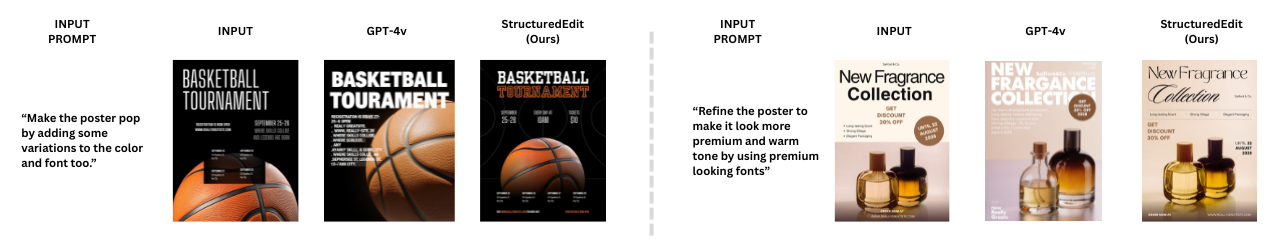}
    \caption{Qualitative comparison: GPT-4V introduces layout violations and typographic errors, while \textsc{StructuredEdit} produces constraint-satisfying, instruction-aligned edits.}
    \label{fig:output}
\end{figure}

We evaluated GPT-4V, standard supervised fine-tuning, and DPP on 25,000 poster designs across A4, A3, letter, 16:9, 1:1, and 9:16 aspect ratios. Evaluation metrics include parameter accuracy (correct extraction and modification of target parameters), constraint satisfaction (percentage of instances that pass all hard constraints), font accuracy (correct font family identification), positioning IoU (spatial alignment of elements), and cost per 1,000 edits. As mentioned in Table~\ref{tab:main}, StructuredEdit improves constraint satisfaction by 37 points over GPT-4V and 21 points over standard SFT at matched inference cost. Matched-element IoU improves from 0.51 to 0.82 and top-1 font accuracy improves from 0.28 to 0.76 on the top-100 Google Fonts. The inference-time rejection rate drops from 35\% (post-hoc validation only) to 12\% with DPP. The outputs from the experiment are showcased in Figure \ref{fig:output}


\noindent \textbf{Ablations.} Removing DPP drops constraint satisfaction by 21 points, confirming that constraint learning, not just fine-tuning data, drives validity gains. Removing hybrid data drops parameter accuracy by 12 points. A symbolic-only variant without differentiable rendering reaches 76\% satisfaction, showing that raster-space feedback captures violations that token-level penalties miss.

\noindent \textbf{User study.} We recruited 35 participants (16 professional designers with 1 to 5+ years' experience, 19 novices). Each completed 6 edit tasks per system in randomized order on a web-based Canva-like interface. Results appear in Table~\ref{tab:userstudy}. Editing time drops 33\%, iterations drop 44\%, and satisfaction and trust both rise significantly. One designer commented that GPT-4V outputs typically required 5 to 10 minutes of
overlap and size fixing before client delivery, while StructuredEdit outputs required minimal cleanup. 

\section{Conclusion}
We presented StructuredEdit, a system for constraint-aware graphic design editing via Differentiable Parameter Propagation. DPP embeds hard constraints into language model training by backpropagating pixel-level violations through differentiable rendering, achieving 89 \% constraint satisfaction versus 52 \% for pixel-based approaches. Hybrid human-AI data generation enables scalable, high-quality training without prohibitive annotation costs. The system demonstrates that parameter-based editing with constraint-aware training outperforms pixel-based generation for precision-critical design tasks, establishing a foundation for controlled creative artifact editing across visual domains.

\bibliographystyle{ACM-Reference-Format}
\bibliography{references}

@inproceedings{Avudaiappan2025PosterDecomp,
  author    = {Avudaiappan, Veeramanohar and Murali, Ritwik},
  title     = {Reconstructing Graphic Design Posters via Visual Decomposition and Semantic Layer Translation},
  booktitle = {ACM SIGGRAPH 2025 Posters},
  series    = {SIGGRAPH Posters '25},
  year      = {2025},
  publisher = {Association for Computing Machinery},
  address   = {New York, NY, USA},
  articleno = {41},
  numpages  = {3},
  location  = {Vancouver, BC, Canada},
  doi       = {10.1145/3721250.3743040},
  url       = {https://doi.org/10.1145/3721250.3743040}
}

@article{Jia2023Cole,
  author  = {Jia, Peidong and Li, Chenxuan and Yuan, Yuhui and Liu, Zeyu and Shen, Yichao and Chen, Bohan and Chen, Xingru and Zheng, Yinglin and Chen, Dong and Li, Ji and Xie, Xiaodong and Zhang, Shanghang and Guo, Baining},
  title   = {{COLE}: A Hierarchical Generation Framework for Multi-Layered and Editable Graphic Design},
  journal = {arXiv preprint arXiv:2311.16974},
  year    = {2023},
  url     = {https://arxiv.org/abs/2311.16974}
}

@inproceedings{Seol2024PosterLlama,
  author    = {Seol, Jaejung and Kim, Seojun and Yoo, Jaejun},
  title     = {{PosterLlama}: Bridging Design Ability of Language Model to Content-Aware Layout Generation},
  booktitle = {Computer Vision -- ECCV 2024},
  series    = {Lecture Notes in Computer Science},
  volume    = {15140},
  pages     = {451--468},
  year      = {2024},
  publisher = {Springer},
  doi       = {10.1007/978-3-031-73007-8_26},
  url       = {https://doi.org/10.1007/978-3-031-73007-8_26}
}

@inproceedings{Yamaguchi2021CanvasVAE,
  author    = {Yamaguchi, Kota},
  title     = {{CanvasVAE}: Learning to Generate Vector Graphic Documents},
  booktitle = {Proceedings of the IEEE/CVF International Conference on Computer Vision (ICCV)},
  year      = {2021},
  pages     = {5481--5489},
  publisher = {IEEE},
  doi       = {10.1109/ICCV48922.2021.00545},
  url       = {https://doi.org/10.1109/ICCV48922.2021.00545}
}

@inproceedings{Brooks2023InstructPix2Pix,
  author    = {Brooks, Tim and Holynski, Aleksander and Efros, Alexei A.},
  title     = {{InstructPix2Pix}: Learning to Follow Image Editing Instructions},
  booktitle = {IEEE/CVF Conference on Computer Vision and Pattern Recognition (CVPR)},
  pages     = {18392--18402},
  year      = {2023},
  publisher = {IEEE},
  doi       = {10.1109/CVPR52729.2023.01764},
  url       = {https://doi.org/10.1109/CVPR52729.2023.01764}
}

@article{Shahgir2025FRB,
  author  = {Shahgir, Haz Sameen and Sayeed, Khondker Salman and Abhik, Roy and Ly, Aaron and Hasan, Akib and Dong, Yue},
  title   = {Texture or Semantics? Vision-Language Models Get Lost in Font Recognition},
  journal = {arXiv preprint arXiv:2503.23768},
  year    = {2025},
  note    = {Published at COLM 2025},
  url     = {https://arxiv.org/abs/2503.23768}
}

@inproceedings{Kikuchi2025,
  title     = {Multimodal Markup Document Models for Graphic Design Completion},
  author    = {Kotaro Kikuchi and Ukyo Honda and Naoto Inoue and Mayu Otani and Edgar Simo-Serra and Kota Yamaguchi},
  booktitle = {ACM International Conference on Multimedia},
  year      = {2025},
  doi       = {10.1145/3746027.3755420}
}

@misc{bryan2013rfc,
  title={RFC 6902: JavaScript Object Notation (JSON) Patch},
  author={Bryan, P and Nottingham, M},
  year={2013},
  publisher={RFC Editor}
}

\end{document}